# Watching a single fluorophore molecule walk into a plasmonic hotspot

*Ling Xin[1,2,‡], Mo Lu[3,‡], Steffen Both[4], Markus Pfeiffer[3], Maximilian J. Urban[1,2], Chao Zhou[1,2], Hao Yan[5], Thomas Weiss[4], Na Liu[1,2]\*, and Klas Lindfors[3]\**

[1] Max Planck Institute for Intelligent Systems, Heisenbergstrasse 3, 70569 Stuttgart, Germany
[2] Kirchhoff Institute for Physics, Heidelberg University, Im Neuenheimer Feld 227, 69120 Heidelberg, Germany
[3] Department of Chemistry, University of Cologne, Luxemburger Straße 116, 50939 Köln, Germany
[4] 4th Physics Institute and Stuttgart Research Center of Photonic Engineering, University of Stuttgart, 70569 Stuttgart, Germany
[5] Department of Chemistry & Biochemistry, Biodesign Institute, Arizona State University, Tempe, AZ 85287-5601, USA

Plasmonic nanoantennas allow for enhancing the spontaneous emission, altering the emission polarization, and shaping the radiation pattern of quantum emitters. A critical challenge for the experimental realizations is positioning a single emitter into the hotspot of a plasmonic antenna with nanoscale accuracy. We demonstrate a dynamic light-matter interaction nanosystem enabled by the DNA origami technique. A single fluorophore molecule can autonomously and unidirectionally walk into the hotspot of a plasmonic nanoantenna along a designated origami track. Successive fluorescence intensity increase and lifetime reduction are in situ monitored using single-molecule fluorescence spectroscopy, while the fluorophore walker gradually approaches and eventually enters the plasmonic hotspot. Our scheme offers a dynamic platform, which can be used to develop functional materials, investigate intriguing light-matter interaction phenomena as well as to serve as prototype system for examining theoretical models.

Plasmonic nanoantennas can convert propagating optical radiation into localized electromagnetic fields in a subwavelength region (i.e. a hotspot) and vice versa.[1,2] Such antennas allow for enhancing the emission as well as shaping the radiation patterns of single emitters including fluorophores,[3–5] fluorescent nanodiamonds,[6,7] semiconductor quantum dots,[8–12] among others. Strong coupling between excitons and plasmons has also been achieved using nanoantennas.[13–15] However, it is technically very challenging to position a single emitter into the hotspot of a plasmonic nanoantenna with nanoscale precision. So far, top-down nanotechniques have often been used for this aim but critical limitations remain.[8,10–12,16,17] For instance, metal nanostructures prepared using top-down nanotechniques are generally polycrystalline, offering poorer optical properties than chemically synthesized nanocrystals.[18] In addition, complex three-dimensional (3D) architectures with engineered hotspots are formidable to achieve. Most critically, the resulting optical properties are intrinsically static, once the devices are fabricated. This impedes a rigorous understanding of light-matter interactions on the nanoscale.

DNA nanotechnology is a viable solution to address the aforementioned challenges. In particular, the DNA origami technique[19,20] allows for creation of 3D DNA architectures with almost arbitrary shapes and full addressability. A variety of nanoscale objects including metal nanoparticles,[21–24] fluorophores,[22,25–27] quantum dots,[22,28,29] carbon nanotubes,[30] proteins,[31,32] etc., can be positioned on DNA origami with nanoscale accuracy.[28,33] It has been demonstrated that hybrid structures containing plasmonic nanoparticles and single emitters assembled on DNA origami can enable enhanced radiative rate of emission,[25,27] increased Förster resonance energy transfer (FRET) rate,[34,35] and single-molecule surface-enhanced Raman scattering.[36,37] However, the published works have been restricted to static assemblies, in which the separations of the plasmonic particles and single emitters remain fixed, allowing for light-matter interactions only under static conditions. Separation tuning has been achieved often by fabricating multiple static samples with different relative positions of the emitters and plasmonic particles. To this end, the remarkable capability of the DNA origami technique has not yet been fully employed.[38–41] Here we demonstrate a dynamic light-matter interaction system, in which a single fluorophore molecule can walk autonomously and unidirectionally into the hotspot of a plasmonic gap antenna along an RNA decorated origami track. Accelerated fluorescence decay on a single structure level is observed in real time, when the fluorophore walker successively approaches and eventually enters the plasmonic hotspot powered by DNAzyme-RNA interactions.[38,39,41] Such autonomous walking is based on a burnt-bridge mechanism, thus giving rise to unidirectional movements without any external intervention.

RESULTS AND DISCUSSION

**Device design.** Figure 1a illustrates the schematic of our nanosystem. A DNA origami template (gray) comprises a bottom platform and a track. Two gold nanoparticles (AuNPs) of 60 nm in diameter are functionalized with DNA strands that are complementary to the capture strands extended from the bottom platform and the track of the origami template. The AuNPs are immobilized on the two sides of the template through DNA hybridization. The track of about 10 nm in width passes through a nominal 15 nm gap in between the two AuNPs. Due to the softness of the structure, we expect small deviations from the design values in the dimensions of the fabricated devices. A fluorophore molecule (red sphere) is attached to a DNAzyme strand (see Fig. 2a, walker), serving as an autonomous fluorophore walker. The overall height of the track and the walker is adjusted such that the fluorophore can eventually walk into the hotspot center of the plasmonic gap antenna. Figure 1b presents a transmission electron microscopy (TEM) image of the assembled structures.

**Fluorescence assay.** We first test the autonomous walking process in the absence of the AuNPs. Figure 2a shows the design principle. Ten DNA footholds as stators (0-9) are arranged in a zigzag fashion along the DNA origami track. The distance between the neighboring stators along the zigzag directions is 5.2 nm, which defines the step size. Eight RNA oligonucleotides (blue) as substrates are positioned at stators 1-8. For testing the autonomous walking process, a quencher molecule (BHQ1, black sphere) is attached to the DNAzyme strand (black), which is positioned at stator 0 by a start strand (purple) and initially blocked from walking by a blocker strand (green). The autonomous walking can halt at stator 9 by a stop strand (red). As shown in Fig. 2b, the fluorophore Cy5 is assembled along the track at stator 4 (red sphere). Upon addition of the DNA trigger strands (brown), the blocker strand (green) is released and the DNAzyme that carries the quencher molecule comes into action. In the presence of divalent metal ions, the DNAzyme can catalyze the cleavage of the individual RNA substrates for unidirectional walking of the walker along the track based on a burnt-bridge mechanism. Figure 2c shows the cleavage of a RNA substrate catalyzed by the 8-17 DNAzyme and its cofactor $Mg^{2+}$.[42,43] The fluorescence intensity of Cy5 is monitored using a fluorescence spectrometer (Jasco FP8500) during the walking process. Time course measurement was used to monitor the sample using an excitation wavelength of 645 nm and the emission was collected at 665 nm. As shown in Fig. 2d, upon addition of the DNA trigger strands, the fluorescence intensity of Cy5 (at stator 4) decreases, reaches a minimum after approximately 15 minutes, and subsequently returns to its initial value after approximately 1 hour. The transient dip implies that the quencher walker departed from stator 0, arrived at stator 4 and then passed it.

**Walk into plasmonic hotspot.** We next investigate the dynamic light-matter interaction process on the level of single devices, while a fluorophore walker autonomously walks into the hotspot of a plasmonic gap antenna (see Fig. 1a), using single molecule fluorescence spectroscopy. In this case, a fluorophore molecule (Atto 647N, absorption and emission maxima at 646 nm and 664 nm, respectively) is attached to the DNAzyme strand. Figure 3a shows an exemplary fluorescence micrograph of the as-fabricated structures. The observed bright spots correspond to immobilized individual structures. All measurements are performed using circularly-polarized incident light. After identifying a region with sufficient devices, the DNA trigger strands are added in the liquid cell to initiate the walking process. The autonomous walking of the fluorophore molecule is monitored by recording fluorescence micrographs and fluorescence decay traces of the selected individual devices at regular time intervals. Figure 3b displays the fluorescence brightness evolution of a representative device as a function of elapsed time, while the fluorophore walker progressively walks towards the hotspot of the plasmonic antenna. As shown in Fig. 3b, the fluorescence brightness increases over time (see also Fig. 4d) and reaches a maximum after approximately 6.5 hours. Meanwhile, the fluorescence decay is successively accelerated (see Figs. 3c and 4d). Fitted monoexponential decay curves for the data measured before walking and at 6.5 hours are shown using red lines in Fig. 3c. The observed brightness increase and lifetime shortening are attributed to the increased excitation rate and total (sum of radiative and non-radiative) decay rate of the fluorophore, when it approaches and eventually enters the hotspot of the plasmonic antenna. We observe that the changes in the fluorescence process for which the data is shown in Figs. 3b and 3c, take place slower than in the experiment of Fig. 2d. In the plasmonic structure, the two large gold nanoparticles fully coated by 15-base-long single-stranded oligonucleotides are located on each side of the walking track, which generates steric hindrance and electrostatic repulsion for the fluorophore walker. These will increase with the walker approaching the hotspot. Additionally, the concentration of the cofactor $Mg^{2+}$ was 20% lower in the experiment with the plasmonic structure (10 mM) than for the fluorescence quenching experiment (12.5 mM). These factors result in the observed slower walking in the plasmonic structure.

To corroborate the experimental observations, finite-element simulations are carried out. Details of the simulations can be found in Supporting Information. The top-view of the simulated structure is depicted in Fig. 4a. In the simulations, the fluorophore (Atto 647N) is represented using an electric point dipole. The excitation wave propagates along the $z$ direction and is circularly polarized within the $xy$ plane as in the experiment. As shown in Fig. 4b, the excitation of the plasmonic antenna results in a strong field enhancement localized in the hotspot. The calculated fluorescence lifetime (top) and intensity (bottom) are shown in Fig. 4c as a function of the relative distance between the dipole and the hotspot center. The orientation of the dipole has critical influences on both the fluorescence lifetime and intensity. For a dipole oriented along the $x$ or $z$ axis, the fluorescence lifetime is only slightly reduced. In contrast, a dipole oriented along the $y$ axis experiences a significant reduction in fluorescence lifetime, resulting from strong interactions with the localized near-fields of the plasmonic antenna. In our structure, a 10.1 Å long linker attaches the fluorophore to the DNAzyme strand. Such long linkers have previously been found to allow the fluorophore to rotate freely.[44] To account for the random dipole orientations in the experiment, the averaged fluorescence lifetime is calculated with the assumption that the fluorophore may rotate in all directions on a timescale faster than its lifetime and can hence be considered as an isotropic emitter.[45,46] For details and further remarks about this approximation, see Supporting Information. The result is presented by the black line in the top panel of Fig. 4c, giving rise to approximately a factor of 7 reduction in lifetime (i.e. ratio of lifetime at the stop and start stator positions). Similarly, the averaged fluorescence intensity is calculated and presented by the black curve in the bottom panel of Fig. 4c. It is evident that the average fluorescence intensity successively increases, as the fluorophore molecule gradually approaches the hotspot. The maximum fluorescence enhancement is approximately a factor of 9. It is noteworthy that for a dipole oriented along the *z* axis, it cannot be excited by the incident wave, and, therefore, the fluorescence intensity remains zero (green curve). We remark that here the simulation results for a gap size of 21 nm are shown. From the design of the DNA origami structure and considering the softness of the structure, we estimate the gap to be between 15 nm and 21 nm. The simulation results for the larger gap size better match the experimental results, so that we choose this value for the simulation. In TEM images, the gap is distorted due to drying of the structure, which makes it impossible to determine the exact value under experimentally relevant conditions.

To quantify the fluorescence modifications, we experimentally characterize a number of devices using single molecule spectroscopy. The top panel of Fig. 4d presents the changes in fluorescence lifetime as a function of the walking time measured from 24 devices. The data of the device in Fig. 3 are replotted using the blue line in the same figure. Here, only the devices, whose lifetime decreases by a factor of at least 1.2 (ratio of lifetime at the start and stop stator positions), are shown. The rest of the devices, which exhibit no observable changes or a complex pattern of changes in fluorescence dynamics are provided in Supporting Information. Their behavior most likely results from missing AuNPs or unsuccessful assembly of the devices (see also Fig. 5). To confirm that the measured signal originates from an Atto 647N molecule, we record fluorescence spectra and investigate the emitters for single-step photobleaching (see Supporting Information). Based on the spectroscopic and time-trace measurements, we conclude that the measured emission originates from single Atto 647N molecules. The error bar in the top panel of Fig. 4d is the average of the standard errors from the fits to the decay traces (see Supporting Information). A monotonous reduction of the fluorescence lifetime is observed, as the walking process continues. The average reduction in lifetime is approximately a factor of 3. The measured changes in fluorescence brightness during the walking process from the same 24 devices are shown in the bottom panel of Fig. 4d. The largest fluorescence enhancement is approximately a factor of 4. Many devices show smaller enhancements or even a decrease of the emission. We speculate that the large spread in the observed changes in brightness is due to sample imperfections resulting from missing particles, inhomogeneous and nonspherical AuNP shapes, as well as defective assemblies (see also Fig. 5). In these cases, the plasmonic antenna might result in partial quenching of the emission. We note that the observed changes in the fluorescence lifetime and intensity are very similar to what has been previously observed in static plasmonic DNA origami devices.[27,47] The error bar in the bottom panel of Fig. 4d is obtained from the average change in brightness for static devices recorded over 17 hours (see Supporting Information). As a control experiment, we characterize individual static devices where the fluorophore is immobilized at stator 0 in the presence of the plasmonic antenna. The control experiment is carried out in the same time span as that for the dynamic devices. No successive fluorescence change in brightness is observable for the static devices.

In addition to actively tracking selected devices during the walk (see Fig. 4d), we determine the fluorescence brightness and fluorescence lifetime for a large number of devices before and after the walking has been completed (elapsed time larger than 20 h). We additionally measure the same data for static devices with the emitter positioned either at the start or stop position. The static devices lack the track but are otherwise identical to the dynamic devices. The results are shown in Fig. 5. For reference DNA origami walker devices, which have the track but are assembled without the gold nanoparticles, we observe a narrow distribution of fluorescence lifetimes and emission intensities. As soon as the emitters are embedded in walker devices containing nanoantennas (Fig. 5, middle panel), the fluorescence lifetime distribution is strongly broadened. We interpret our results to be a consequence of deviations in the geometry of the device from the design. When the emitters are in the hotspot of the antenna (dynamic devices after the walk or static devices with the emitter positioned in the hotspot, Fig. 5, in the rightmost panel), the center of the distribution of the fluorescence lifetime shifts to smaller values due to the increase in the total decay rate. At the same time, the emission brightness distribution is broadened towards large values. In the rightmost panel of Fig. 5 we show with red solid line the theoretically calculated emission intensity as a function of lifetime using the same parameters as for Fig. 4c. Each point of the simulated line corresponds to a different position along the track. We observe that the experimental data points lie below the simulated result. This illustrates that small defects in the structure or a halt of the walker decrease the obtained enhancement of the emission. We finally note that the results presented in Fig. 5 are in good agreement with the data from the dynamic experiments in Fig. 4 and that the obtained values for the fluorescence enhancement are similar to those obtained in previous work on static plasmonic DNA origami structures.[27,47]

CONCLUSIONS

In summary, we have demonstrated dynamic control of light-matter interactions using the DNA origami technique. Our system is powered by DNAzyme-RNA reactions, giving rise to autonomous and unidirectional motion of a fluorophore walker. The fluorescence decay of the fluorophore is accelerated when the fluorophore walker gradually approaches and eventually enters the hotspot of a plasmonic nanoantenna. Our approach can be used to achieve complex dynamic plasmonic architectures with tailored functionalities. For example, incorporation of multiple fluorophores or semiconductor quantum dots within a plasmonic structure could allow for dynamic switching of the interactions between the emitters as well as their coupling to the far field. Our experimental approach further allows for optically monitoring the autonomous motion of a single DNA machine with high temporal and spatial resolution.

METHODS

**DNA origami assembly.** DNA origami was assembled from 10 nM scaffold strands and 100 nM of each set of staple strands (10-fold excess) in 1 × Tris-acetate buffer with 12.5 mM magnesium acetate (pH 7.6), using a 20 h annealing program (85°C 5 min, 70°C-61°C -1°C/min, 60°C-50°C -1°C/60 min, 49°C-25°C -1°C/20 min, and 23°C hold).

**DNA machine loading on DNA origami template.** For devices in fluorescence assay (Figure 2d), the quencher-walker (walker-BHQ1) was assembled with blocker and start strand to form a complex Q with a 1:1:1 molar ratio. Then, the complex Q, substrates, and stop strand were mixed with Cy5-labeled DNA origami assembly with a ratio of 5:40:5:1. Two fluorescent complexes were first assembled from the fluorophore-walker (walker-Atto 647N): the walker was assembled with blocker and start strand to form a complex A with a 1:1:1 molar ratio, and with stop strand to form a complex B. Then, for dynamic devices, the complex A, substrates, and stop strand were mixed with DNA origami assembly with a ratio of 5:40:5:1; while for static devices, the corresponding fluorescent complex was mixed with DNA origami assembly with a ratio of 5:1, respectively. The mixtures were incubated overnight at room temperature. All devices were purified by 0.7% agarose gel electrophoresis in the DNA origami annealing buffer for 4 h at 8 V/cm in a gel box immersed in an ice-water bath.

**DNA functionalization of the AuNPs.** Functionalization of the 60 nm AuNPs with 5' end thiolated DNA (HS-T15, 5'-HS-TTT TTT TTT TTT TTT-3') was carried out following the procedure: Thiolated DNA strands were incubated with TCEP [tris(2-carboxyethyl)phosphine] with a ratio of 1:200 for 1 h. Before functionalization, the AuNPs were spun down and then resuspended with modification buffer (0.5 × TBE buffer containing 0.02% of SDS, pH 8.0). AuNPs (0.1 nM, 0.5 mL) were mixed with 80 μM thiolated DNA strands (5 μL). We used a well-established salt aging procedure,[48] and the final concentration of NaCl reached 0.7 M. The AuNPs functionalized with DNA were then purified by 5 times centrifugation at a rate of 3,000×g, 5 min for each time. The supernatant was carefully removed and the remaining AuNPs were mixed with the purified Atto 647N-labeled dynamic/static devices with a ratio of 8:1 in modification buffer with 10 mM $MgCl_2$. The mixture was incubated overnight at room temperature.

**Immobilization of devices.** Atto 647N-labeled devices modified with AuNPs were purified by 0.7% agarose gel electrophoresis in DNA origami annealing buffer for 4 h at 8 V/cm in a gel box immersed in an ice-water bath. The purified samples were immobilized on a BSA-biotin-neutravidin surface in a commercial LabTek chamber using the procedure of Ref. [49]. 10 μL solution of the gel purified devices was diluted with 300 μL of PBS, 10 mM $MgCl_2$ solution. Each LabTek chamber was then incubated with a specific sample solution (dynamic/static) for 3 hours. The chambers were then washed three times with 650 μL of PBS, 10 mM $MgCl_2$. Each chamber was kept in 300 μL of PBS, 10 mM $MgCl_2$ solution at 4°C.

**Single-molecule spectroscopy.** Immobilized DNA devices in LabTek chambers were characterized in an in-house developed fluorescence microscope. Light from a 635 nm wavelength pulsed laser diode with a repetition rate of 40 MHz was focused to a diffraction-limited spot on the sample using an oil immersion microscope objective with numerical aperture (NA) 1.49. The emitted fluorescence photons were collected with the same objective, separated from the excitation light using dielectric filters, and directed to a fast single-photon counting module or spectrometer. The fluorescence decay traces were recorded using time-correlated single-photon counting by connecting the timing output of the detector to a time tagger. The fluorescence decay traces were analyzed by fitting a monoexponential decay curve to the data (DecayFit - Fluorescence Decay Analysis Software 1.4, FluorTools, www.fluortools.com).

**Finite-element simulations.** The numerical simulations were performed using the commercial software COMSOL Multiphysics. The AuNPs were modeled as gold spheres (permittivity data from Ref. [50]) with a diameter of 60 nm, surrounded by water (refractive index 1.332). The relative change in the intensity was calculated following Refs. [51,52], and the fluorescence lifetime was obtained as the inverse of the total decay rate[53]. Details on the calculations are provided in the Supporting Information. The intrinsic quantum yield $q_0 = 0.65$ of Atto 647N was taken from the specifications provided by the supplier (http://www.atto-tec.com), while the intrinsic lifetime $\tau_0 = 4.09$ ns and the emission spectrum were determined from our reference measurements (dye attached to DNA, but without AuNPs). As mentioned in the main text, the orientation average in Fig. 4c was calculated under the assumption that, in our system, Atto 647N can be approximated as an isotropic emitter. The corresponding equations, as well as a detailed discussion that supports this approximation, are given in the Supporting Information.

FIGURES

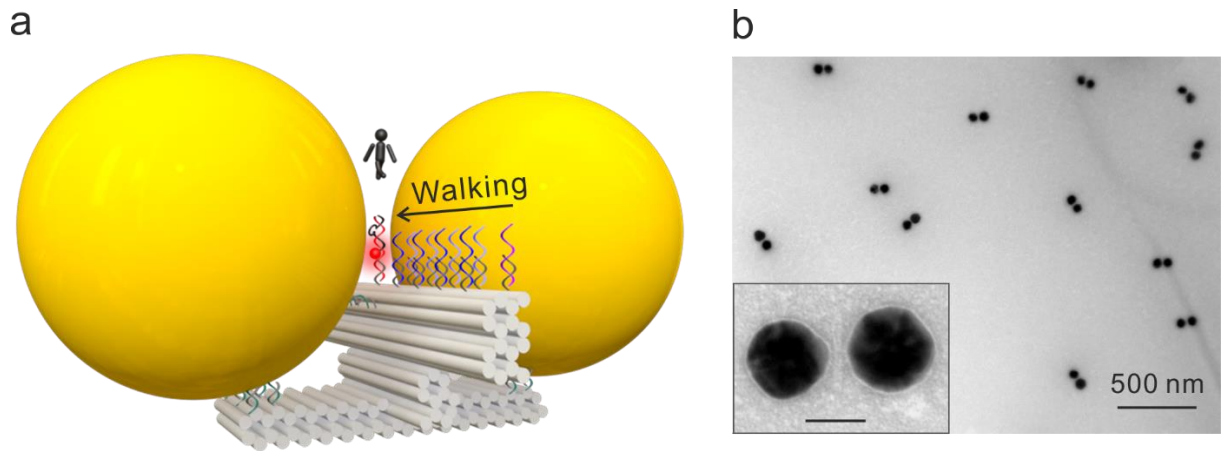

**Figure 1.** (a) Schematic illustration of the dynamic light-matter interaction system. It contains two gold nanoparticles (60 nm) assembled on a 3D DNA origami template, which consists of a platform and a track bundle. A fluorophore molecule (red sphere) carried by a DNAzyme strand can autonomously walk into the plasmonic hotspot along the origami track. (b) TEM image of the assembled structures and enlarged view (inset) of a representative structure. Scale bar: 50 nm.

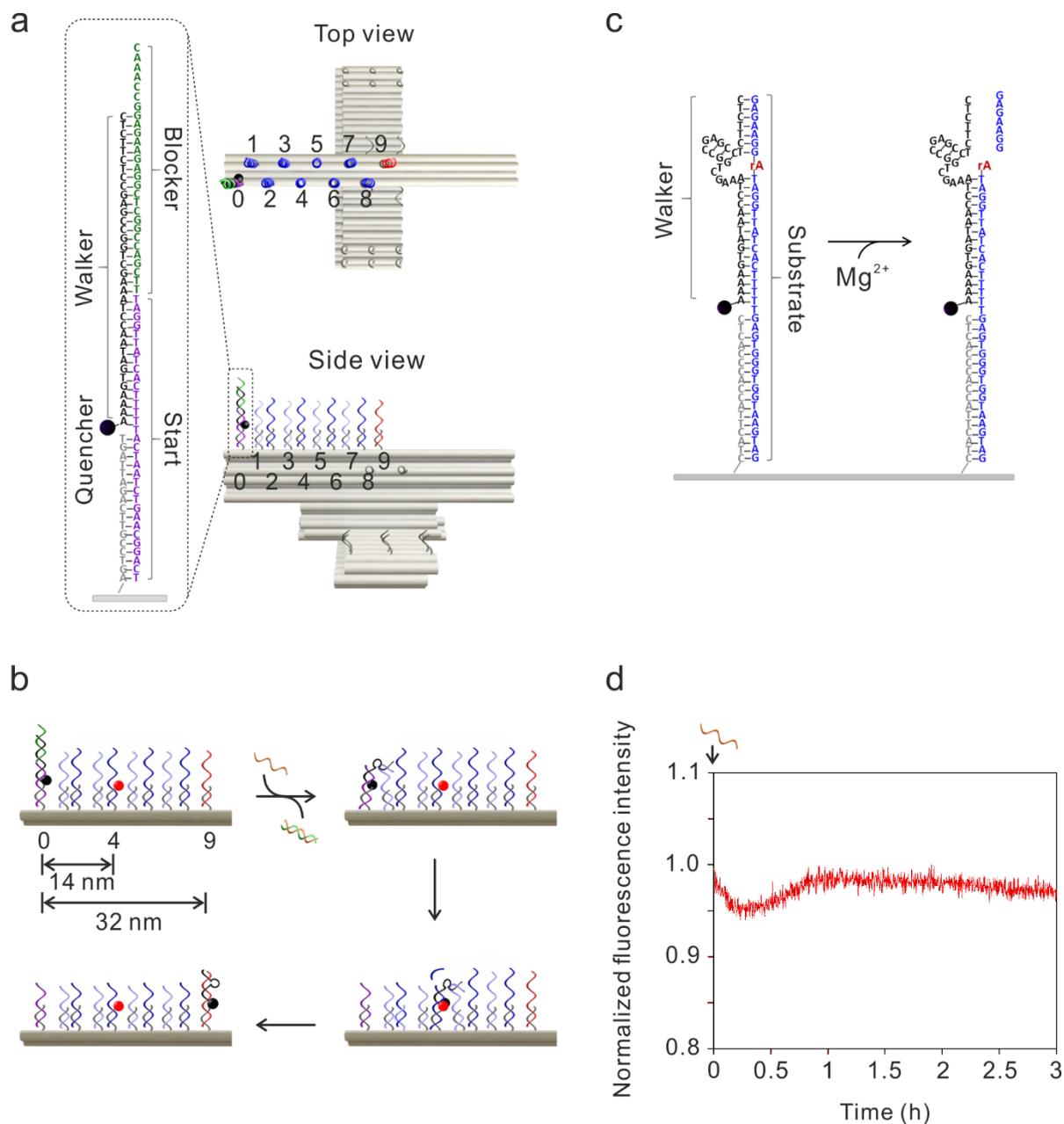

**Figure 2.** (a) Top- and side-view of the walking track, which comprises ten stators (0-9) arranged in a zigzag fashion along the origami bundle. A quencher molecule (BHQ1, black sphere) is attached to the DNAzyme strand (black), which is positioned at stator 0 by a start strand (purple) and initially blocked from walking by a blocker strand (green). The autonomous walking can halt at stator 9 by a stop strand (red). (b) The fluorophore Cy5 is assembled at stator 4 (red sphere) along the origami track. Upon addition of the DNA trigger strands (brown), the blocker strand (green) is released and the 8-17 DNAzyme can start autonomous walking. (c) In the presence of $Mg^{2+}$ ions (12.5 mM in this experiment), the DNAzyme cleaves its substrate at the RNA base (rA), resulting in two shorter strands (7 and 16 bases in length, respectively). Dissociation of the upper part allows the walker to interact with the subsequent substrate on the track. (d) The fluorescence of Cy5 is momentarily quenched, while the quencher walker walks along the track and passes the fluorophore.

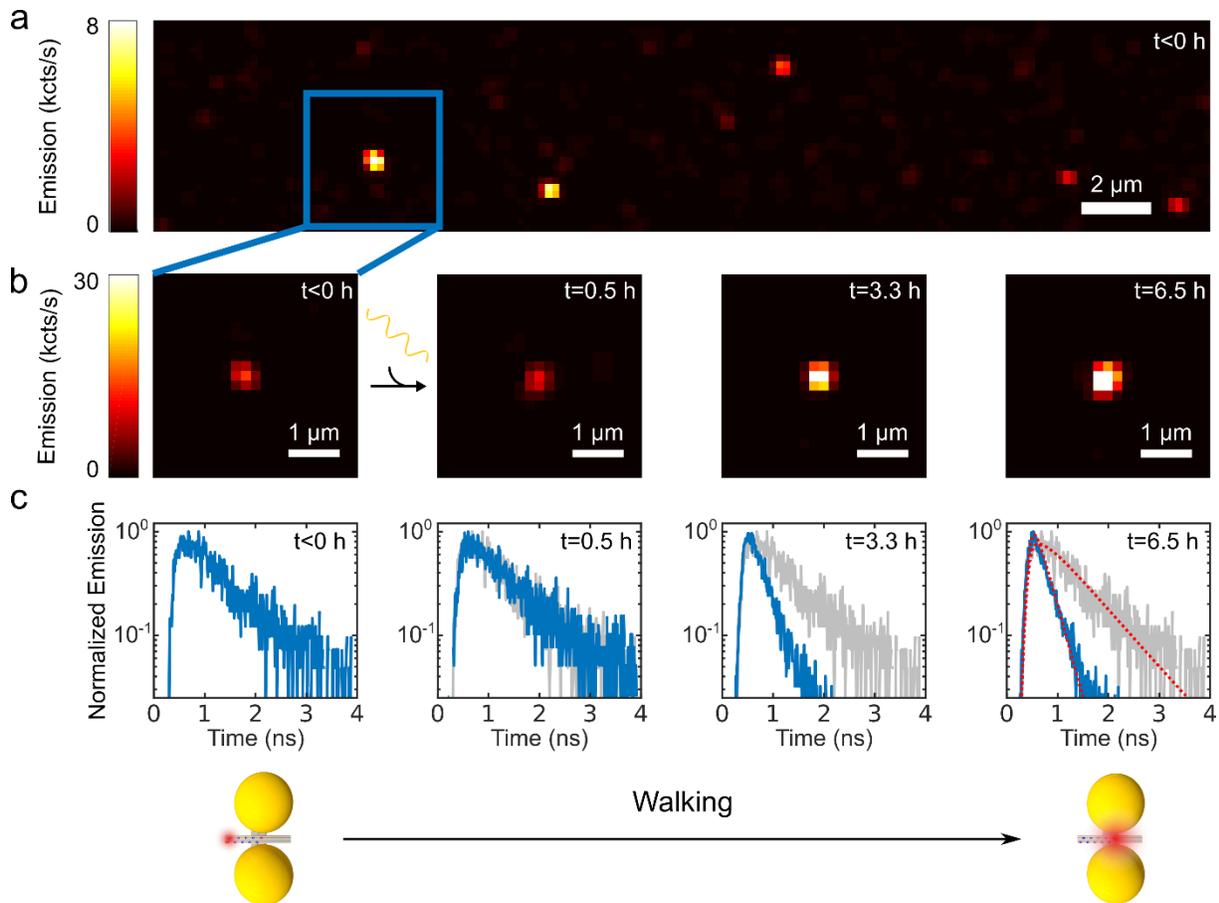

**Figure 3.** Fluorescence microscopy of single fluorophore walker devices. (a) Fluorescence micrograph of the immobilized devices before walking (t<0 h) at a peak excitation intensity of approximately 300 W/cm$^2$. (b) Increase in fluorescence brightness of a single device [marked with blue square in (a)] as a function of elapsed time. The micrographs are recorded before walking and at times of 0.5 h, 3.3 h, and 6.5 h after walking is started. (c) Changes in fluorescence dynamics for the same device as in (b) as a function of elapsed time. The decay curve measured before walking (gray solid line) is shown as a reference in the panels obtained at later times. The fits used to extract the fluorescence lifetimes are shown for the last time point (t=6.5 h) with red dotted lines.

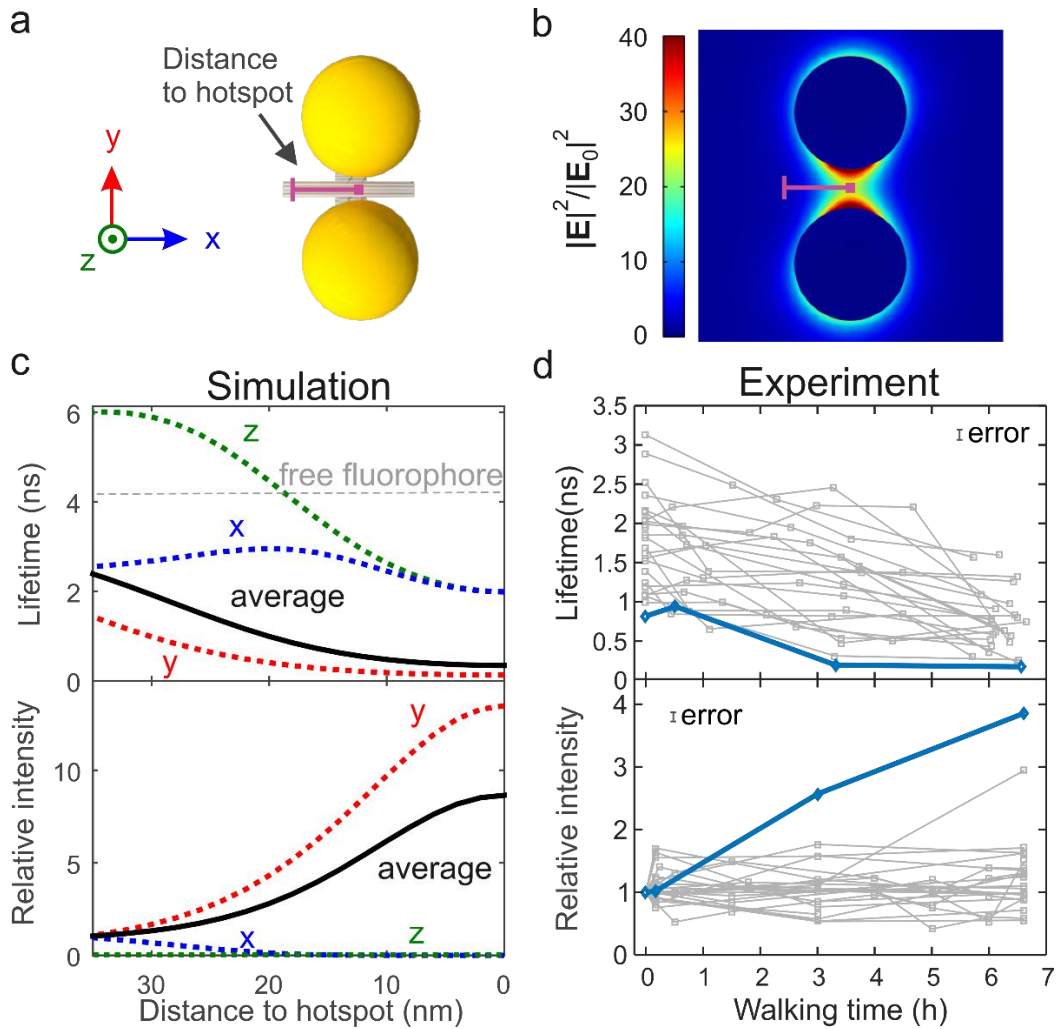

**Figure 4.** Quantification of the changes in fluorescence. (a) Schematic illustration of the track and definition of the coordinate system used in the simulations. The purple line indicates the distance to the hotspot along the track. The excitation wave propagates along the $z$ direction (out of plane) and is circularly polarized within the $xy$ plane. The gap of the gold nanoparticle dimer antenna is assumed to be 21 nm (see main text). (b) Relative intensity enhancement for illumination with a plane wave at a wavelength of 635 nm, calculated with the finite-element method. (c) Calculated lifetime (top) and relative intensities (bottom) at different positions along the track. The intrinsic lifetime of the free fluorophore in the absence of gold nanoparticles was measured to be 4.09 ns from DNA walker devices (gray dashed line). The dashed lines denote the results for a fluorophore with a fixed orientation along the $x$ (blue), $y$ (red), and $z$ (green) axis, respectively. The solid black line represents the average value. The relative fluorescence intensities are normalized to the intensity at the start stator position of the walker. (d) Experimentally measured decrease in fluorescence lifetime (top) and increase in fluorescence brightness (bottom) as a function of elapsed time. Each trace corresponds to a single device. The device presented in Figs. 3b and 3c is shown in blue curves.

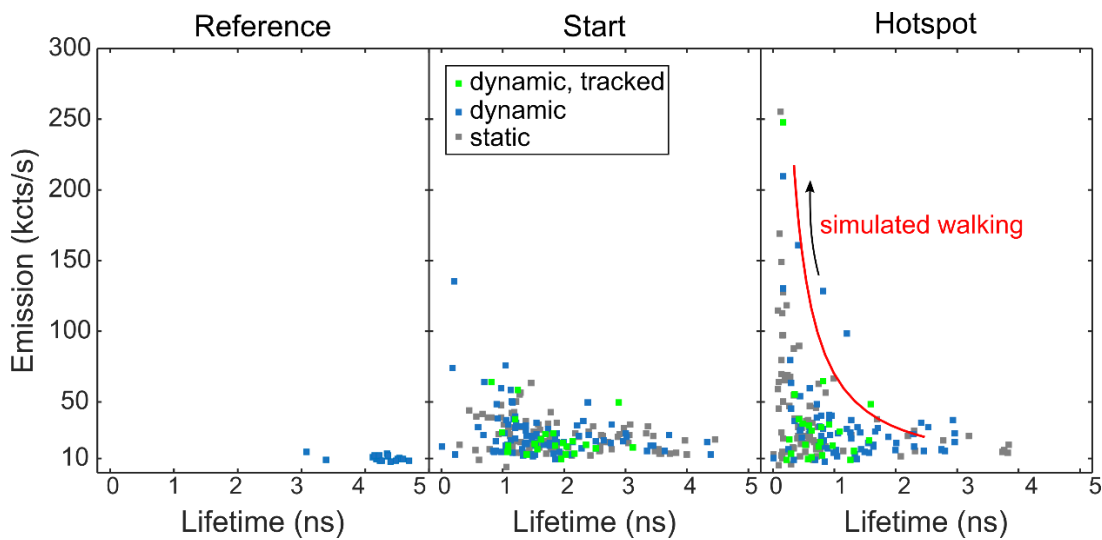

**Figure 5.** Emission intensity as a function of fluorescence lifetime. (left) For DNA walker devices lacking gold nanoparticles (reference), (middle) for devices with the walker at the start position (start), and (right) after walking (hotspot). The data points correspond to: the dynamic devices that were actively tracked during the walk (data in Fig. 4d) (green), dynamic devices before or after walking (blue), and static devices (gray). The red solid line shows the simulated emission intensity as a function of lifetime when the emitter moves from the start position (longest lifetime) to the hotspot (shortest lifetime).

ASSOCIATED CONTENT

**Supporting Information**. Additional details on sample preparation, single molecule spectroscopy, and finite-element simulations.

AUTHOR INFORMATION

**Corresponding Author**

*E-mail (N.L.): naliu@is.mpg.de.
*E-mail (K.L.): klas.lindfors@uni-koeln.de.

**Author Contributions**

N.L. conceived the concept. L.X., M.U., and C.Z. designed the DNA origami nanostructures. L.X. and M.U. performed the DNA assembly experiments. L.X. did the track design and ensemble fluorescence walker measurements. M.L., M.P., and K.L. performed the single-molecule fluorescence experiments and data analysis. S.B. and T.W. carried out the theoretical calculations. H.Y. made helpful suggestions. The manuscript was written through contributions of all authors. All authors have given approval to the final version of the manuscript.

‡These authors contributed equally.

**Notes**

The authors declare no competing financial interest.

ACKNOWLEDGMENT

This work was supported by a grant from the Volkswagen foundation. N.L. was supported by the Sofja Kovalevskaja grant from the Alexander von Humboldt-Foundation and the European Research Council (ERC Dynamic Nano) grant. M.U. acknowledges the financial support by the Carl-Zeiss-Stiftung. M.L. and M.P. were supported by a UoC Postdoc grant. Funding from the University of Cologne through the Institutional Strategy of the University of Cologne within the German Excellence Initiative (QM2) is acknowledged.

105005.